\documentstyle[preprint,revtex]{aps}
\tightenlines
\begin{document}
\draft
\preprint{ULLFT--1/93\\}
\preprint{February 1993}

\begin{title}
{\bf The Process of Generation of Mass, The Higgs Boson,
and Dark Matter}
\end{title}

\author{Vicente Delgado}

\begin{instit}
{Departamento de F\'{\i}sica Fundamental y Experimental,\\
Universidad de La Laguna, E-38203 La Laguna, Tenerife, Spain.}
\end{instit}

\begin{abstract}
A dynamical mechanism of symmetry breaking in which gauge and
matter fields play an active role is proposed. It basically
represents a covariant generalization of the mechanism
responsible for superconductivity, and provides a {\em natural}
mechanism of generation of mass which is not in conflict
with the present value of the cosmological constant.
When applied to
SU(2)$\times$U(1) leads to exactly the same physics (Lagrangian
density) as the Standard Model but modifying {\em only} the
Higgs sector. It also predicts the
appearance over all space of a classical scalar field as well
as the existence of density fluctuations.
According to it, space
would be filled with a macroscopically large number of Higgs
bosons which now appear as light,
stable scalar particles {\em decoupled} from ordinary matter and
radiation. Therefore they would play the same role as
the Cooper pairs in superconductivity and would
be a natural candidate for dark matter.
\end{abstract}
\pacs{PACS numbers: 12.50.L, 14.80.G, 95.35, 98.80.C}

\narrowtext

Spontaneous symmetry breaking is one of the basic underlying
ideas of both Unified Gauge Theories and Inflationary
Cosmology. This phenomenon occurs when the ground state of
a system displays less symmetry
than that of the corresponding Hamiltonian, and is related
to the concept of vacuum and the process of mass generation.
Its importance lies mainly on the fact that the only
renormalizable gauge theories with massive vector
bosons are gauge theories with spontaneous symmetry
breaking \cite{tHooft}.
In modern Unified Theories massive intermediate bosons are
properly introduced by breaking the symmetry by means of
the Higgs mechanism. It basically consist in the introduction
of a complex scalar field $\phi(x)$ subject to an
effective potential of the form
\begin{equation}
V(\phi)=\mu^2\phi^+\phi + \lambda(\phi^+\phi)^2
\end{equation}
with $\mu^2 < 0$. The necessity of introducing a negative
mass squared for the scalar field is the price one has to pay
in order to generate
a nonzero stable configuration $\langle\phi \rangle_0$.

In this paper I propose a different mechanism where
spontaneous symmetry breaking has its origin in the dynamics
of the scalar field $\phi$ interacting with gauge and matter
fields, which now play an active role. In fact, symmetry
breaks once a particular solution of the
dynamical equations is chosen among a (one-parameter) family of
possible solutions.

In what follows I will consider in detail the case of an abelian
U(1) theory, which basically represents a relativistic
generalization of the mechanism responsible for
superconductivity.
This analogy with superconductivity proves to be very useful in
order to gain valuable
insights about the process of symmetry breaking.
Specific application to the Standard Model (SM) will
be considered below.

The U(1) gauge invariant Lagrangian density for a complex
scalar field $\phi(x)$ reads
\begin{equation}
{\cal L}=(D_\mu{\phi})^+({D^\mu{\phi}}) - \mu^2 \phi^+ \phi
- \frac{1}{4} {\rm F}_{\mu\nu} {\rm F}^{\mu\nu}
\end{equation}
where now $\mu^2 > 0$ is the mass of the scalar field and the
${\rm F}_{\mu\nu}$ tensor is given by
${\rm F}_{\mu\nu}(x) = \partial_\mu B_\nu - \partial_\nu B_\mu$.

It is convenient parametrize $\phi(x)$ as
\begin{equation}\label{phipar}
\phi(x)= {1 \over {\sqrt 2}} \rho(x) \exp i\xi(x)
\end{equation}
where $\rho(x)$ and $\xi(x)$ are real scalar fields.
Then, in the unitary gauge, where
$\phi(x) \rightarrow \rho(x)/\sqrt{2}$,
the Lagrangian density can be rewritten as
\FL
\begin{equation}
{\cal L} = \frac{1}{2}(\partial_\mu \rho)(\partial^\mu \rho) +
\frac{1}{2} g^2 A_\mu A^\mu \rho^2 - \frac{1}{2} {\mu^2}\rho^2
- \frac{1}{4} {\rm F}_{\mu\nu} {\rm F}^{\mu\nu}
\end{equation}
with
$A_\mu(x)= B_\mu(x) - (1/g)\partial_\mu \xi(x)$.

The  Euler-Lagrange  equations  of  motion  which  describe  the
coupled
dynamics of the fields, are
\begin{eqnarray}
\label{ero}
&&\partial_\mu \partial^\mu \rho(x)= - \mu^2 \rho(x) +
g^2 A^2(x) \rho(x)
\\
\label{efaa}
&&\partial_\alpha \partial^\alpha A^\nu -\partial^\nu
(\partial_\beta A^\beta) = -g^2 A^\nu \rho^2
\end{eqnarray}
where the source of the gauge field,
$J^\mu(x)=A^\mu(x) \rho^2(x)$,
is a conserved current.

In what follows I will show that the equations of motion
(\ref{ero})--(\ref{efaa}) admit a family of solutions which
verify
\begin{equation}\label{sol}
A^\mu(x)= {J^\mu(x) \over \rho^2(x)}
\end{equation}
{\em where} $J^\mu(x)$ is not an independent function of
$\rho(x)$, and a subset of these solutions leads to a theory
with broken symmetry and massive gauge bosons.

By substituting Eq.(\ref{sol}) into Eq.(\ref{ero})
we obtain
\begin{equation}\label{rove}
\partial_\alpha \partial^\alpha \rho(x) =-{\partial \over
\partial \rho}
V_{\rm eff}(x)
\end{equation}
where
\begin{equation}
V_{\rm eff}(x)=\frac{1}{2} \mu^2 \rho^2(x) + g^2
{J^2(x)\over 2\rho^2(x)}
\end{equation}
In field theories the vacuum $\vert 0 \rangle$ is defined as
the ground
state of the theory. The property of translational invariance
that this state must possess requires vacuum expectation values
(VEVs) of physical magnitudes to be constants independent of
space-time coordinates. Then we have
\begin{equation}
\begin{array}{l}
\langle 0 \vert \rho(x) \vert 0 \rangle = \langle \rho
\rangle_0 = v\\ { }\\
\langle 0 \vert A_\mu(x) A^\mu(x) \vert 0 \rangle =
\langle A_\mu A^\mu
\rangle_0\\ { }\\
\langle 0 \vert J^2(x) \vert 0 \rangle = \langle J^2
\rangle_0 = l^2
\end{array}
\end{equation}
so that the vacuum $\vert 0 \rangle$ must verify
\begin{equation}\label{evac}
\left.{{\partial \over \partial \rho} V_{\rm eff}(x)}\right
\vert_{\langle { }
\rangle} =\mu^2
\langle \rho \rangle_0 -
g^2 {\langle J^2 \rangle_0 \over \langle \rho^3 {\rangle_0}} = 0
\end{equation}
which is nothing but the equation obtained by proyecting
Eq.(\ref{rove}) onto the vacuum.
Different values of $l^2$ characterize different
possible solutions and therefore different possible
vacuum states.
This means that the properties of the vacuum depend on the
current squared
$J^2(x)$. Thus gauge (and matter) fields play an
active role in selecting the ground
state $\vert 0 \rangle$. Note, however, that due to the fact
that $l^2$ is a constant it is not possible to smoothly pass
from one to another solution.
We have indeed a one-parameter family of stable solutions,
and the U(1)
symmetry breaks once a particular one is taken.

The subset of solutions we are interested in are those with
$l^2 > 0$, because
as can be seen  from Eq.(\ref{evac}) for $l^2 \neq 0$ it follows
that $\langle \rho \rangle_0 = v \neq 0 $. In this case we have
\begin{equation}
\label{vlu}
\mu^2\langle \rho \rangle_0\langle \rho^3 \rangle_0 =g^2 l^2
\end{equation}
Furthermore, Eq.(\ref{sol}) leads to
\begin{equation}\label{el2}
l^2=\langle \rho^4 \rangle_0\langle A_\mu A^\mu \rangle_0
\end{equation}
Therefore the solutions we are interested in satisfy
\begin{equation}\label{laa}
\mu^2\langle \rho \rangle_0 \langle \rho^3 \rangle_0=
g^2\langle \rho^4 \rangle_0 \langle A_\mu A^\mu \rangle_0
\end{equation}
Notice that a similar equation, which relates the mass of the
scalar field $\rho(x)$ to the vacuum fluctuations of the
gauge field, can be directly obtained from the initial
equations of motion. Indeed by proyecting
Eq.(\ref{ero}) onto the vacuum one finds that solutions with
$\langle \rho \rangle_0 \neq 0$
verify
\begin{equation}\label{ule}
\mu^2=g^2 \langle A_\nu A^\nu \rangle_0
\end{equation}
This equation represents a necessary condition for
a solution with $\langle \rho \rangle_0 \neq 0$ can be taken.
It basically states that in order for the process to take place,
$\mu$ particles should be created from energy fluctuations
of gauge fields in the ground state.

By comparing Eqs.(\ref{laa}) and (\ref{ule}) one expects
\begin{equation}\label{vn}
\langle \rho^n \rangle_0 \simeq \langle \rho \rangle_0^n = v^n
\end{equation}
so that we can rewrite Eq.(\ref{vlu}) as
\begin{equation}\label{v4l2}
v^4={g^2 \over \mu^2} l^2
\end{equation}
On the other hand, by expanding $V_{\rm eff}(x)$ about the
vacuum one obtains
\begin{eqnarray}\label{vefant}
V_{\rm eff}(x) =&& \left. V_{\rm eff} \right
\vert_{\langle { } \rangle}
+ \frac {1}{2!} \left. {\partial^2 \over \partial
\rho^2} V_{\rm eff}\right \vert_{\langle { } \rangle}
(\rho(x) - v)^2
\nonumber\\
&&+ \frac{1}{3!}
\left. {\partial^3 \over \partial \rho^3} V_{\rm eff}
\right \vert_{\langle { } \rangle} (\rho(x) - v)^3 + \ldots
\end{eqnarray}
where terms proportional to $(J^2(x)-l^2)$, which have no
influence on the dynamics of $\rho(x)$, have been omitted.
Substitution of Eq.(\ref{vefant}) into Eq.(\ref{rove})
leads to
\FL
\begin{equation}
\partial_\alpha \partial^\alpha \rho(x) + m^2 (\rho(x)-v)=
\lambda (\rho(x)-v)^2 + \ldots
\end{equation}
where
\begin{eqnarray}
m^2=&&\left.{\partial^2 \over \partial \rho^2} V_{\rm eff}
\right \vert_{\langle { } \rangle} =\mu^2 +
3 {g^2 l^2 \over v^4}=4\mu^2\\
\lambda=&& -\frac{1}{2}\left. {\partial^3 \over \partial
\rho^3} V_{\rm eff}
\right \vert_{\langle { } \rangle} =
6 {g^2 l^2 \over v^5}={6 \mu^2 \over v}
\end{eqnarray}
Then defining a real scalar field with vanishing VEV
\begin{equation}\label{dfeta}
\eta(x)=\rho(x)-v
\end{equation}
the equations of motion finally read
\begin{eqnarray}
\label{effr}
&&(\partial_\alpha \partial^\alpha  + m^2) \eta(x)= \lambda
\eta^2(x)
+ \ldots \\
\label{eaf}
&&(\partial_\alpha \partial^\alpha + m^2_A )A^\nu -\partial^\nu
(\partial_\beta A^\beta) = j^\nu(x)
\end{eqnarray}
where
\begin{equation}\label{jcor}
j^\nu(x) = -(2g m_A \eta(x)+g^2\eta^2(x)) A^\nu(x)
\end{equation}
{}From these equations we see that for
a solution with $l^2>0$, the scalar field develops a
nonvanishing VEV
\begin{equation}
\phi(x)= {1\over \sqrt {2}}(v+\eta(x)) \exp i\xi(x)
\end{equation}
and the gauge field $A_\mu(x)$ acquires a mass
\begin{equation}
\label{mbos}
m^2_A=g^2 v^2
\end{equation}
Futhermore, according to Eq.(\ref{effr}) the field $\eta(x)$
(as well as $\rho(x)=v+\eta(x)$)
{\em decouples}. Thus, they become no
affected by the process of interaction with gauge fields. Indeed
its  dynamics  becomes  governed  by  the  effective  Lagrangian
density
\FL
\begin{equation}\label{Leff}
{\cal L}_{\rm eff}=\frac {1}{2} (\partial_\mu \eta)
(\partial^\mu \eta)
-\left( \left. V_{\rm eff} \right \vert_{\langle { } \rangle} +
\frac{1}{2} m^2 \eta^2 - {\lambda \over 3} \eta^3 +
\ldots \right)
\end{equation}
Quantum fields exhibit this kind of behaviour in the limiting
case where
a macroscopically large number of quanta appear in a coherent
state, and in such a situation they behave as ordinary
classical fields.

In particular, given a solution with $l^2>0$,
the field $\rho(x)$, which describes $\mu$ particles, behaves
as a classical scalar field which fluctuates about $\rho(x)=v$
with a mass
$m=2\mu$. Then one expects that a macroscopic
physical meaning should be possible to be given to $\rho(x)$.
Indeed, it can be interpreted as describing a Bose condensate
of $\mu$
particles with vanishing momenta and a number density given by
\begin{equation}
\label{densi}
n(x)=\frac {1}{2} m \rho^2(x)
\end{equation}
With this interpretation the minimun of the potential energy
density
corresponds, as expected, to the internal energy of a Bose
condensate. In fact, it can be written as
\begin{equation}
\left. V_{\rm eff} \right \vert_{\langle { } \rangle}=
\mu^2 v^2= \mu n_0
\end{equation}
with $n_0=\langle n(x) \rangle_0$ being the mean particle
density. On the other hand, according to Eqs.(\ref{mbos}) and
(\ref{densi}), the gauge boson acquires a mass
\begin{equation}
m_A^2 = {g^2 n_0 \over \mu}
\end{equation}
which also agrees with what one would expect.
In fact, $m_A^{-1}$ coincides with the London penetration depth
for a condensate of $\mu$ particles with number density $n_0$.
Futhermore, for small fluctuations about the equilibrium
configuration,
$\eta(x) \ll v$, Eq.(\ref{densi}) leads to
\begin{equation}
\eta(x) \simeq {1 \over mv} (n(x)-n_0)
\end{equation}
which shows that the classical scalar field $\eta(x)$ can be
regarded in turn
as describing density
fluctuations about the mean value $n_0$. Incidentally, note that
the mass of this field corresponds to the energy necessary to
create a pair of $\mu$ particles.

Thus Eqs.(\ref{effr})--(\ref{eaf}) would describe a gauge field
which evolves
in a classical scalar medium, acquiring mass and producing
in turn density fluctuations in this medium.
These results are consistent with the interpretation that for
a solution
with $l^2 > 0$, a phase transition takes place and a
Bose condensate
of $\mu$ particles is formed.

Once the phase transition takes place, the
dynamics  of  $A_\mu(x)$  becomes  governed  by  the  Lagrangian
density
\begin{equation}
{\cal L'} = - \frac{1}{4} {\rm F}_{\mu\nu} {\rm F}^{\mu\nu} +
\frac{1}{2} m_A^2 A_\mu A^\mu + {\cal {L'}_{\rm I}}
\end{equation}
where
\begin{equation}
{\cal {L'}_{\rm I}} = gm_A\eta A_\mu A^\mu
+\frac {1}{2} g^2 \eta^2 A_\mu A^\mu
\end{equation}
describes the interaction of the quantum field $A_\mu(x)$ with
the external
classical field $\eta(x)$ governed by the Lagrangian density
${\cal L_{\rm eff}}$ given by Eq.(\ref{Leff}).

In order to get a better understanding on the mechanism
which causes the decoupling of the scalar field, let us
rewrite Eq.(\ref{effr}) retaining now the term in
$(J^2(x)-l^2)$ which was neglected before
\begin{equation}
\label{misma}
(\partial_\alpha \partial^\alpha  + m^2) \eta(x)=
{3 m^2 \over 2v} \eta^2(x) +{g^2 \over v^3} (J^2(x)-l^2)
+ \ldots \\
\end{equation}
This equation explicitly shows that the decoupling occurs
when $v \rightarrow \infty$, which according to
Eq.(\ref {densi}) means $n_0 \rightarrow \infty$. Thus
the decoupling is caused by the presence of a
macroscopically large number of quanta of the scalar field,
which then behaves as a classical field.

Note finally that the classical external medium  seems  to  play
the
role of
a regulator. In fact, if one tries to decouple the {\em massive}
gauge field from the scalar medium by taking $g  \rightarrow  0$
in
Eqs.(\ref{eaf})--(\ref{jcor}), then Eq.(\ref{ule}) would imply
$\langle A_\mu A^\mu \rangle_0 \rightarrow \infty$.

The above mechanism can also be applied to the SU(2)$\times$U(1)
Standard Model in a
straigthforward  manner.  However,  in  this  case  it  is  also
possible
to break the symmetry starting from a massless scalar field,
which seems a
more  natural  alternative.  In  what  follows  I  will  briefly
consider
this case. A detailed derivation will be given elsewhere
\cite{VDB2}.

We start from the Lagrangian density of the SM with the {\em only}
modification that now the Higgs potential {\em vanishes}. Then,
in terms of eigenstates of the mass operator the Euler-Lagrange
equation of
motion for the massless scalar field, in the unitary gauge has
the form
\FL
\begin{eqnarray}
\label{eqrsm}
\partial_\mu \partial^\mu \rho(x) =&& \frac {1}{4} \left( 2g^2
W_\mu^- W^{\mu+}
+(g^2+{g'}^2) Z_\mu Z^\mu \right)\rho(x) \nonumber \\
&& - \sum_f {\lambda_f \over \sqrt{2}}
\overline{\psi}_f \psi_f
\end{eqnarray}
where $W_\mu^\pm$ and $Z_\mu$ are the mass eigenstates of the
weak vector
bosons, $\psi_f$ denotes a Dirac spinor, and the sum runs over
massive fermions
corresponding to the three families.
As for the equations of motion of the gauge and matter fields,
they are exactly the same as those of the Standard Model and
are not explicitly shown.

In order to get a theory with broken symmetry and massive gauge
and fermion
fields, we look for solutions of the equations of motion
with the requirement that the sources of the 'free'
gauge and matter
fields are not independent functions of $\rho(x)$. In this
case Eq.(\ref{eqrsm}) takes the form of Eq.(\ref{rove}),
where now $V_{\rm eff}(x)$ is given by
\begin{equation}
V_{\rm eff}(x)=\frac{1}{8} {J_G^2(x) \over \rho^2(x)} -
{J_F^2(x)\over \rho(x)}
\end{equation}
and here $J_G^2(x)$ and $J_F^2(x)$ play the role of
gauge and fermion currents squared, respectively.

On the other hand, by proyecting onto the vacuum
$\vert 0 \rangle$ we find that this state must verify
\begin{equation}
\label{vasm}
\left.{{\partial \over \partial \rho} V_{\rm eff}(x)}\right
\vert_{\langle { }
\rangle} = - \frac {1}{4} {\langle J_G^2 \rangle_0 \over
\langle \rho^3 {\rangle_0}} + {\langle J_F^2 \rangle_0 \over
\langle \rho^2 {\rangle_0}} = 0
\end{equation}
It is not difficult to see (by proyecting
Eq.(\ref{eqrsm}) onto $\vert 0 \rangle$ and comparing
with Eq.(\ref{vasm})), that
these solutions lead to a nonvanishing VEV $\langle \rho
\rangle_0 = v \neq 0$
only if
$\langle \rho^4 \rangle_0=\langle \rho^3 \rangle_0 \langle \rho
\rangle_0$,
so that we are again led to look for solutions satisfying
Eq.(\ref{vn}). Then, from Eq.(\ref{vasm}) we have
\begin{equation}
v=  \frac {1}{4} {\langle J_G^2 \rangle_0 \over \langle J_F^2
\rangle_0}
\end{equation}
which shows that a particular solution with
$\langle J_G^2 \rangle_0, \langle J_F^2 \rangle_0 >0$ breaks
the symmetry.

By expanding $V_{\rm eff}(x)$ about the vacuum, as before, one
finally obtains
\begin{equation}
\label{this}
(\partial_\alpha \partial^\alpha  + m_H^2) \eta(x)= \lambda
\eta^2(x)
+ \ldots
\end{equation}
where $\eta(x)=\rho(x)-v$, and
\begin{eqnarray}
\label{hola}
m_H^2 &&= 2{g^2\over 4} \langle W_\mu^- W^{\mu+}\rangle_0
+{(g^2+{g'}^2)\over 4} \langle Z_\mu Z^\mu \rangle_0
\nonumber \\
&&=  \sum_f {\lambda_f \over \sqrt{2} \, v}
\langle\overline{\psi}_f \psi_f\rangle_0 \\
\lambda &&={3m_H^2 \over v}
\end{eqnarray}
Substitution of $\rho(x)=v+\eta(x)$ into the equations of motion
of the gauge and matter fields then leads to exactly the same
equations for the corresponding massive quantum fields as the SM.
This means that the dynamics of the gauge and matter fields
would be governed by an effective Lagrangian density which
coincides with that of the SM but interpreting now the field
$\eta(x)$ as an external classical field. Therefore the present
mechanism would lead to exactly the same physics as the SM
{\em except} for the Higgs sector.

On the other hand, when $\rho(x)$ evolves towards its stable
configuration in $V_{\rm eff}(x)$, it acquires a  nonvanishing
mass $m_H$
from vacuum fluctuations of quantum fields. In this process a
phase transition takes place and it develops
a nonzero VEV $\langle \rho \rangle_0= v$, reflecting the
appearance
of a macroscopically large number of quanta of the field.
These quanta
are Higgs bosons created from the vacuum energy.
The process gives rise to the formation of a Bose condensate
of Higgs particles, and as a
consequence $\rho(x)=v+\eta(x)$ behaves as a decoupled
classical field. Then, as long as the Standard Model remains
valid everywhere, there must exist a macroscopic classical
scalar field over all space. This can give us a relation
between microphysics and the macroscopic world. According
to this mechanism starting from a microscopic system
interacting with a scalar field, as a result of a phase
transition a macroscopic classical
medium can arise where quantum fields evolve. Gauge and fermion
fields would acquire mass as a consequence of the inertia they
manifest
evolving through this classical medium. In fact, this is a
natural mechanism for particles to get mass. Motion of charged
particles
inside a cristal lattice, through a plasma, or through a strong
electromagnetic field are some examples where particles respond
with a larger inertia.

According to the present mechanism
space would be filled with a macroscopically large number of
Higgs bosons
created in the early universe from the vacuum energy.
Being stable, scalar
particles decoupled from ordinary matter and radiation,
the Higgs bosons
only would be detectable by gravitation and they would
be a natural candidate for dark matter.

On the other hand, it can be shown that the vacuum energy
density $\rho_V$,
given by the minimun of $V_{\rm eff}(x)$, takes the form
\begin{equation}
\label{denva}
\rho_V= \left. V_{\rm eff} \right \vert_{\langle { } \rangle}=
- \frac {1}{2} m_H^2 v^2
\end{equation}
Thus, the experimental value of $\rho_V$
can be used in order to estimate an upper bound to the mass
of the Higgs
boson. From Eq.(\ref{denva}), one has
\begin{equation}
m_H^2= {2\vert \rho_V \vert \over v^2}
\end{equation}
so that in order to be consistent with the present value of
the cosmological
constant, $\Lambda=8\pi G \vert \rho_V \vert \leq 10^{-84 }
{\rm GeV}^2$, it
is only required
\begin{equation}
m_H \leq 10^{-26} {\rm GeV}
\end{equation}
It should be notice that this bound is not an specific feature
of the mechanism of symmetry breaking proposed in this paper.
In fact, if one demands the Higgs mechanism to be consistent
with the present value of the cosmological constant then one
obtains the same result \cite{Dreit}.
However, the Higgs mechanism cannot consistently account for
both the scale of electroweak symmetry breaking (which
requires $v \simeq 250$ GeV) and such a small mass for the
Higgs boson (or what is the same, such a small value for the
cosmological constant). The reason is that the Higgs mechanism
leads to $m_H^2=2 \lambda_c v^2$ ($\lambda_c$ being the quartic
coupling constant) so that one expects $m_H$ to be of
order $O(v)$
(unless $\lambda_c$ takes an unnatural small value).
According to the present mechanism, however, the scale of the
electroweak symmetry breaking and the mass of the Higgs
boson are related {\em only} through the value of the
vacuum energy density, Eq.(\ref{denva}), and the surprisingly
small value of $m_H$ would simply reflect the
small value of the cosmological constant.

In order to try to understand how such a small mass could be
understood in the context of the Standard Model let us take
a closer look at the way in which masses generate.
Gauge and matter fields acquire mass
evolving inside a dense medium described by a
field $\rho(x)=v+\eta(x)$ with a nonvanishing mean value.
This is basically a classical mechanism (apart from the well
known examples of superconductivity or motion of classical
particles inside dense media, consider, for instance, plasma
oscillations or the screening mechanism of Debye-H\"uckel)
and the corresponding masses are of order $O(v)$.
However, the way in which the Higgs boson acquires mass is
completly different. As Eq.(\ref{hola}) shows its mass
appears as a purely quantum effect. In fact, taking into
account that the dynamics of the Higgs boson,
Eq.(\ref{this}), becomes governed by
the Lagrangian density
\begin{equation}\label{Leffot}
{\cal L}_{\rm eff}=\frac {1}{2} (\partial_\mu \eta)
(\partial^\mu \eta)-
\frac{1}{2} m_H^2 \eta^2 + {m_H^2 \over v} \eta^3 +
\ldots
\end{equation}
we see that $m_H$ explicitly breaks the
invariance of ${\cal L}_{\rm eff}$ under dilatations
(scale invariance). Then we would have a partially
conserved dilatation current and the mass of the
Higgs would reflect the explicit breaking of this
symmetry.

On the other hand, as stated before, once the phase transition
has taken place
the classical field $\rho(x)$ can be regarded as describing a
Bose condensate of Higgs bosons with a number density
\begin{equation}
\label{otra}
n_H(x)= \frac {1}{2} m_H \rho^2(x)
\end{equation}
Thus the density of dark matter  $\rho_{dm}$ would be given by
\begin{equation}
\label{sea}
\rho_{dm}= m_H \langle n_H \rangle_0 = \frac{1}{2} m_H^2 v^2
\end{equation}
so that by using $\rho_{dm}\approx 10^{-29} {\rm gr/cm}^3$ one
obtains a number density of Higgs bosons
\begin{equation}
\langle n_H \rangle_0 \approx 10^{21} {\rm part/cm^3}
\end{equation}
Notice that Eq.(\ref{sea}) imposes a non trivial
relationship among the experimental quantities $v$, $\Lambda$,
and $\rho_{dm}$, so that, even if one does not identify
the Higgs bosons with the dark matter one would obtain
a matter density of Higgs bosons comparable with the
density of (dark) matter $\rho_{dm}$.

On the other hand, Eq.(\ref{denva}) can be rewritten in terms
of $\rho_{dm}$ in the form
\begin{equation}
\label{esta}
\rho_V= - \rho_{dm}
\end{equation}
This equation basically states that the dark matter
(Higgs bosons)
has been created from the vacuum energy density when the
scalar field evolved to the minimun
of the effective potential.
Therefore this mechanism predicts $\vert \rho_V \vert=
\rho_{dm} $. This
could explain why apparently the vacuum energy density and the
matter density
are of the same order of magnitude \cite{Wein}.

In the ground state of the theory (i.e. the vacuum) the Higgs
bosons are
uniformely distributed with a number density
$\langle n_H \rangle_0$.
According to Eq.(\ref{esta}), in this state they would
shield or screen the (negative) energy
density of the vacuum
\begin{equation}
\rho_V + \rho_{dm} = 0
\end{equation}
This situation is quite similar to that encountered in an
ionic plasma.
In fact, from Eq.(\ref{otra}) follows that small deviations
from the
ground state give rise to the appearance of density fluctuations
\begin{equation}
\eta(x) \simeq {1 \over m_H v} (n_H(x)-\langle n_H\rangle_0)
\end{equation}
Then, the classical scalar field $\eta(x)$, governed by
Eq.(\ref{this}),
would describe small fluctuations
about a uniform  distribution of (dark) matter, and the
universe would be
quite similar to a superconductor.

\end{document}